# REPLY to Comments on "Giant Dielectric Response in the One-Dimensional Charge-Ordered Semiconductor $(NbSe_4)_3I$" and "Colossal Magnetocapacitance and Colossal Magnetoresistance in $HgCr_2S_4$" (cond-mat/0607500)

P. Lunkenheimer[1], J. Hemberger[1], V. Tsurkan[1,2], D. Starešinić[3], K. Biljaković[3], and A. Loidl[1]

[1]*Experimental Physics V, Center for Electronic Correlations and Magnetism, University of Augsburg, D-86135 Augsburg, Germany*
[2]*Institute of Applied Physics, Academy of Sciences of Moldova, MD-2028 Chisinau, R. Moldova*
[3]*Institute of Physics, P.O.B. 304, HR-10001 Zagreb, Croatia*

In the present work we reply to the Comment by Catalan and Scott [1] on two of our papers [2,3]. This Comment has been rejected from publication in Physical Review Letters and, hence, our Reply is based on the cond-mat version [1]. We agree with the statement of Catalan and Scott that in dielectrics that are poor insulators, exotic dielectric phenomena can easily arise from non-intrinsic reasons. Typical examples were already pointed out by us, e.g., in [4,5], the most recent being the giant dielectric constant values in $CaCu_3Ti_4O_{12}$ [5]. As acknowledged by Catalan and Scott, in the present two cases we experimentally could exclude any electrode effects [2,3,6]. Instead they propose alternative mechanisms, for the case of $HgCr_2S_4$ (HCS) speculating about a non-stoichiometric surface layer resulting from the chlorine [7] transport agent used for crystal preparation. Similar critics could apply for the magnetoelectric behavior in $CdCr_2S_4$ (CCS) [8], which was prepared in a similar way. However, this scenario can be excluded because for our dielectric measurements, platelets were prepared by considerably polishing down the original crystals thus removing any possible non-stoichiometric layers. In addition, electron probe microanalysis revealed an almost ideal stoichiometry of the crystals [8]. Nevertheless, while external boundary contributions (from electrodes or non-stoichiometric layers) do not play any role, we have to repeat our statement in [3] that an internal boundary mechanism (of so far unknown origin) cannot be fully excluded.

Concerning the unpublished results of Cheong *et al.* (ref. 7 in [1]), the appearance of the magnetoelectric effect in the thio-spinels is highly sensitive to details of sample preparation and impurity concentration [6,9]. For example, annealing in vacuum of single-crystalline CCS samples suppresses the magnetoelectric effect and the relaxation features [6,9]. The effects also are absent in pure polycrystals of CCS [6], but are recovered after tempering or in In-doped polycrystals. Obviously, the origin of the magnetoelectric effect and the role of stoichiometry and defects in the thio-spinels still is an open question. The number of vacancies is closely correlated with internal stresses and as-grown crystals show minimal stress by accommodating the defect concentration [10]. The frozen-phonon calculations by Fennie and Rabe [11] reveal that the infrared-active phonon modes are rather stable. But it is clear from recent experimental [12] and theoretical work [13] that a classical soft-mode concept does not hold in strongly coupled multiferroics. More exotic mechanisms as considered in our original publications [3,8] or a temperature dependence of the effective plasma frequencies, indicative for charge transfer and concomitant increasing covalency, must be considered.

In order to model our results on the temperature-dependent dielectric constant of $(NbSe_4)_3I$ [2], Catalan and Scott had to assume two regions with similar conductivity at room temperature and a higher activation energy of the thinner region, which leads to a much smaller conductivity at low temperatures. However, in this case the overall conductivity (not shown in [1]) looks quite different than our experimental results (compare Fig. 1 and Fig. 1(b) of [2]). In addition, at low temperatures the dc conductivity of the sample would be dominated by the regions with higher activation energy, leading to a behavior indicated by the solid line, in clear contrast to our dc results (cf. line in Fig. 1(b) of [2]).

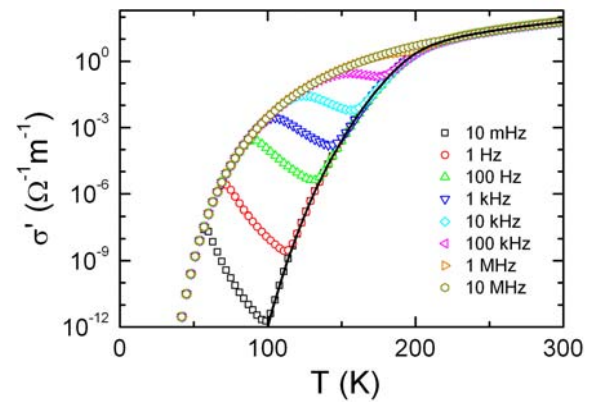

FIG. 1. Conductivity calculated according to the model of Catalan and Scott [1]. The line indicates the expected dc conductivity.

In addition, it seems unlikely that "breaks in the conduction chains" [1] due to microcracks, etc., should lead to nearly identical behavior in samples with very different geometries as in the present case [2]. Finally, as pointed out in [2], $(NbSe_4)_3I$, which is very closely related to the charge density wave (CDW) systems $(TaSe_4)_2I$ and $(NbSe_4)_{10/3}I$, behaves dielectrically very similar as CDW systems. The intrinsic nature of this typical dielectric response of CDWs is well settled and there is a well-founded theoretical explanation for it in terms of screening effects of the pinned CDW [14].